\documentclass[conference,10pt]{IEEEtran}
\IEEEoverridecommandlockouts
\usepackage{cite}
\usepackage{amsmath,amssymb,amsfonts}
\usepackage{graphicx}
\usepackage{textcomp}
\usepackage{xcolor}
\usepackage{here}
\usepackage{listings}
\usepackage{url}
\usepackage{hyperref}
\usepackage{xspace}
\usepackage{multicol}
\usepackage{multirow}
\usepackage{booktabs}
\usepackage{pythonhighlight}
\usepackage{enumitem}[shortlabels]
\usepackage{subcaption}
\usepackage{pgfplots}
\usepackage{amsmath}
\usepackage{algpseudocode}
\usepackage{algorithm}
\usetikzlibrary{patterns}

\algnewcommand\algorithmicforeach{\textbf{for each}}
\algdef{S}[FOR]{ForEach}[1]{\algorithmicforeach\ #1\ \algorithmicdo}

\def\BibTeX{{\rm B\kern-.05em{\sc i\kern-.025em b}\kern-.08em
    T\kern-.1667em\lower.7ex\hbox{E}\kern-.125emX}}

\iffalse
    \newcommand{\an}[1]{\textcolor{cyan}{}}
    \newcommand{\aidan}[1]{\textcolor{red}{}}
    \newcommand{\clg}[1]{\textcolor{orange}{}}
    \newcommand{\ruben}[1]{\textcolor{magenta}{}}
    \newcommand{\daniel}[1]{\textcolor{blue}{}}
    \newcommand{\vmm}[1]{\textcolor{green}{}}
\else
    \newcommand{\an}[1]{\textcolor{cyan}{\bf\small [AN: #1]}}
    \newcommand{\aidan}[1]{\textcolor{red}{\bf\small [ADN: #1]}}
    \newcommand{\clg}[1]{\textcolor{orange}{\bf\small [CLG: #1]}}
    \newcommand{\ruben}[1]{\textcolor{magenta}{\bf\small [RM: #1]}}
    \newcommand{\daniel}[1]{\textcolor{blue}{\bf\small [DR: #1]}}
    \newcommand{\vmm}[1]{\textcolor{green}{\bf\small [VMM: #1]}}
\fi

\newcommand{\method}{SOAR\xspace}

\newcommand{\eg}{\textit{e.g., \xspace}}
\newcommand{\ie}{\textit{i.e., \xspace}}
\newcommand{\tf}{TensorFlow\xspace}
\newcommand{\torch}{PyTorch\xspace}
\newcommand{\dplyr}{dplyr\xspace}
\newcommand{\pandas}{pandas\xspace}
\newcommand\lt[1]{{\lstinline+#1+}} 
\newcommand{\reb}[1]{\textcolor{black}{#1}}

\definecolor{dkgreen}{rgb}{0,0.5,0}
\definecolor{dkred}{rgb}{0.5,0,0}
\definecolor{gray}{rgb}{0.5,0.5,0.5}
\begin{document}

\title{\method: A Synthesis Approach\\for Data Science API Refactoring
%\thanks{Identify applicable funding agency here, post anonymous review.}
}

\author{\IEEEauthorblockN{Ansong Ni\IEEEauthorrefmark{1} \thanks{\IEEEauthorrefmark{1}Both authors contributed equally to this work.}}
\IEEEauthorblockA{%\textit{dept. name of organization (of Aff.)} \\
\textit{Yale University}\\
New Haven, USA \\
ansong.ni@yale.edu}

\and

\IEEEauthorblockN{Daniel Ramos\IEEEauthorrefmark{1}}
\IEEEauthorblockA{\textit{INESC-ID/IST, U. Lisboa,} Portugal \\
\textit{Carnegie Mellon University,} USA\\
danielrr@cmu.edu}

\and

\IEEEauthorblockN{Aidan Yang}
\IEEEauthorblockA{%\textit{School of Computer Science} \\
\textit{Queen’s University}\\
Kingston, Canada  \\
a.yang@queensu.ca}

\and

\IEEEauthorblockN{Inês Lynce}
\IEEEauthorblockA{%\textit{School of Computer Science} \\
\textit{INESC-ID/IST, U. Lisboa}\\
Lisboa, Portugal  \\
ines.lynce@tecnico.ulisboa.pt}

\and

\IEEEauthorblockN{Vasco Manquinho}
\IEEEauthorblockA{%\textit{School of Computer Science} \\
\textit{INESC-ID/IST, U. Lisboa}\\
Lisboa, Portugal  \\
vasco.manquinho@tecnico.ulisboa.pt}

\and

\IEEEauthorblockN{Ruben Martins}
\IEEEauthorblockA{\textit{School of Computer Science} \\
\textit{Carnegie Mellon University}\\
Pittsburgh, USA  \\
rubenm@cs.cmu.edu}

\and

\IEEEauthorblockN{Claire Le Goues}
\IEEEauthorblockA{\textit{School of Computer Science} \\
\textit{Carnegie Mellon University}\\
Pittsburgh, USA  \\
clegoues@cs.cmu.edu}

}

\maketitle

\begin{abstract}
With the growth of the open-source data science community, both the number of data science libraries and the number of versions for the same library are increasing rapidly.
To match the evolving APIs from those libraries, open-source organizations often have to exert manual effort to refactor the APIs used in the code base. 
Moreover, due to the abundance of similar open-source libraries, data scientists working on a certain application may have an abundance of libraries to choose, maintain and migrate between. 
The manual refactoring between APIs is a tedious and error-prone task. Although recent research efforts were made on performing automatic API refactoring between different languages, previous work relies on statistical learning with collected pairwise training data for the API matching and migration. Using large statistical data for refactoring is not ideal because such training data will not be available for a new library or a new version of the same library.
We introduce Synthesis for Open-Source API Refactoring (\method), a novel technique that requires no training data to achieve API migration and refactoring. 
\method relies only on the documentation that is readily available at the release of the library to learn API  representations and mapping between libraries. Using program synthesis, \method automatically computes the correct configuration of arguments to the APIs and any glue code required to invoke those APIs.
\method also uses the interpreter's error messages when running refactored code to generate logical constraints that can be used to prune the search space.
Our empirical evaluation shows that \method can successfully refactor 80\% of our benchmarks corresponding to deep learning models with up to 44 layers with an average run time of 97.23 seconds, and 90\% of the data wrangling benchmarks with an average run time of 17.31~seconds.
\end{abstract}


\section{Introduction}
\label{sec:Intro}
\iffalse

CLG notes to self on related work:

Historical and Impact Analysis of API Breaking Changes: A Large-Scale Study

Statistical Learning Approach for Mining API Usage Mappings for Code Migration

D. Konstantopoulos, J. Marien, M. Pinkerton, and E. Braude, “Best
principles in the design of shared software,” in 33rd International
Computer Software and Applications Conference (COMPSAC), 2009,
pp. 287–292.

S. Moser and O. Nierstrasz, “The effect of object-oriented frameworks
on developer productivity,” Computer, vol. 29, no. 9, pp. 45–51, 1996

\fi

Modern software development makes heavy use of libraries, frameworks, and
associated \emph{application programming interfaces} (APIs). Libraries provide
modular functionality intended for reuse, with prescribing a particular architecture~\cite{jaspan2009checking},
and their widespread use has important productivity
advantages~\cite{de2004good}. The API for a library defines the interface, or
contract, between the (hidden) library implementation of a piece of library functionality, and its client component~\cite{maalej2013patterns}.
Good API selection and maintenance is a key component of modern software
engineering~\cite{de2009roles}. 

Although ideally API selection and usage
could be stable over the course of a software project's lifetime, there are many
practical reasons that client code must update the way it uses a given API, or
even which API/library it uses for a given set of functionality. Broadly,
software may evolve because of a change in the code, the documentation, its properties,
or the customer-experienced
functionality~\cite{chapin2001types}. The APIs used by the software can become
invalid or inapplicable as the software evolves. APIs themselves may become
deprecated or obsolete~\cite{perkins2005automatically}. As a
result, to maintain and optimize software that depends on APIs, developers often
have to refactor APIs between different versions or to another API (\ie 
\textit{API migration}) altogether.

API migration is a form of software refactoring, a critical software engineering
activity that is largely
performed manually~\cite{kim2012field} and is tedious and often error-prone~\cite{kim2017data}.
Migration can be difficult even when migrating between two closely-related
APIs that nominally provide the same functionality. For example, consider
increasingly popular data science and deep learning libraries, such as
TensorFlow~\cite{abadi2016tensorflow}, PyTorch~\cite{paszke2017automatic}, and
Numpy~\cite{walt2011numpy}. Moving between two such libraries often requires
significant manual labor as well as domain- and library-specific knowledge (we
illustrate with an example in Section~\ref{sec:Motivating}); worse, APIs can
change, and outdated historical knowledge can exacerbate these challenges.

Fortunately, many popular APIs possess key properties that can inform an
automated approach to support migration or evolution. First, open-source APIs are often
reasonably well-documented~\cite{zhong2009inferring}. The quality, quantity, and structure of
that documentation can vary widely~\cite{uddin2015api}, but as code intended to be called
and reused by unrelated client applications, documentation is often key to
successful API uptake~\cite{uddin2015api}. Second, unsuccessful API methods
often raise exceptions with informative error messages, that developers can use
to access stack traces and information that can help them modify a
program~\cite{hartmann2010would}. We observe that data science API error
messages are particularly useful as these error messages often identify how the
input data relates to the raised exception. Take for example  ``\textit{Error in fit[5, 100]:
subscript out of bounds}'', which is an error message describing an index overflow. From the example error message, we know that either 5 or 100 is out of bounds for the input matrix. Third, although multiple APIs may vary in concrete implementation details, it is often possible to discretely map between pieces of functionality between source and target APIs (by virtue of solving the same general sets of problems). 

%\an{the intro looks too long to me, ideally we want to get to this on the first
%page} 
% CLG: agreed generally, and technically we're there, so I'll declare vistory
% for now. I bet we could shorten the abstract, too.  
We propose \method (Synthesis for Open-source API Refactoring), a novel approach
that combines natural language processing (NLP) with program
synthesis~\cite{gulwani2017program} to automatically migrate/refactor between
APIs. We focus our approach and evaluation on deep learning and data science
APIs. Since there are
many APIs targeting these domains, changes and new releases are introduced rapidly (as one
example, TensorFlow had 26 releases in 2019 alone), 
and switching between them is common and often tricky~\cite{guo2019empirical}.  Moreover,
data scientists and other users of such libraries have broad backgrounds and
are not always classically trained programmers, and thus could particularly
benefit from tool support to assist them in these tasks~\cite{kim2017data}.
However, we believe the approach will generalize to other APIs with similar
properties (see detailed discussion in Section~\ref{sec:Evaluation}).

Given a program that uses a given source API, \method's central proposition is to use
NLP models learned over available API documentation and error messages to inform program
synthesis to replace all source API calls with corresponding calls taken from
the target API.  \method starts by using existing documentation for the source
and target libraries to build an \emph{API matching model}, which finds likely
replacement calls for each API call in the source program. 

However, simply finding the right function in the corresponding target API is
not enough. The new function must be called with the correct arguments,
and function specifications may vary between libraries.  
\method uses \emph{
  program synthesis} to construct the full target method call in a way that
replicates the original source behavior.

This synthesis step may be further informed by specifications inferred, again,
from the API documentation.

During the program synthesis enumeration procedure,
a potential migrated call may throw an error when
tested. In these situations, \method uses an 
\emph{error message understanding} model that again uses NLP 
techniques to analyze error messages and generate logical 
constraints to prune the search space of the synthesis task.

To the best of our knowledge, \method is the first refactoring tool that
incorporates program synthesis and machine learning tools for refactoring, and
 is a significant improvement over the prior state of the art. 
\method maps programs between different APIs using only readily-available
documentation.  It does not require
manual migration mappings~\cite{meqdadi2019bug} or a
history of previous migrations or refactorings in other software
projects~\cite{savga2008comeback,nguyen2014statistical,bui2019sar,gu2017deepam}.  
Indeed, \method does not require training data at all, and is thus
applicable for migrations to a new library or newer version
of the same library shortly after release. We demonstrate that \method is
versatile in Section~\ref{sec:Evaluation}, using it to migrate between two 
 deep learning libraries (\ie \tf to \torch) in the same programming language
(\ie Python) and between two data manipulation libraries (\ie \dplyr to \pandas) in two
 different programming languages (\ie R and Python).  Prior techniques either
 specialize exclusively in supporting cross-language migration (\eg
 StaMiner~\cite{nguyen2014statistical}), or do 
not support it at all. 
Because \method uses synthesis, when it succeeds the produced code is
guaranteed to compile and pass existing test cases for the original source code.

In summary, our main contributions are:
\begin{enumerate}
	
	\item
	We propose \method, a novel approach based on NLP and synthesis for
	automatic API refactoring, focusing on (but not limited to) deep
	learning and data science tasks.
    \item 
    \method requires no training data, and its output is guaranteed to compile and pass existing test cases. Instead of using training data from prior programs, \method leverages API documentation and program error messages to generate logical constraints to prune the program enumeration search space.
	\item 
	We evaluate \method on two library migration tasks (\ie \tf to \torch
	and \dplyr to \pandas) to demonstrate its effectiveness. 
	Our results show that \method can successfully migrate
	80\% of neural network programs composed by 3 to 44 layers in with an 
	average time of 97.23 seconds. And for \dplyr to \pandas migration, 90\% of benchmarks are solved on average in 17.31 seconds.
    \item With ablation studies, we also evaluate how each part of
	\method impacts its performance. We show that the use of specifications from API documents and learning from error messages are largely helpful for the synthesis process. We also show how different API matching methods perform on the two migration tasks. %\an{I added this, please check}
	\item
	We release the \method implementation for the two migration tasks mentioned above. We also release the documentation and benchmark tests we use in this work to facilitate future research on this direction. 
	
\end{enumerate}

The remainder of this paper is organized as follows: 
Section~\ref{sec:Motivating} presents a motivating example that illustrates the challenges of manual API refactoring. In section~\ref{sec:Algorithm}, we describe our approach to automatic API migration. Section~\ref{sec:Evaluation} presents our empirical evaluation and analysis of results. Next, we discuss our current approach and limitations in section~\ref{sec:Limitations}. Finally, we conclude with an overview of related work in section~\ref{sec:Background} and conclusions in section~\ref{sec:Conclusions}.

\section{Motivating Example}
\label{sec:Motivating}
\begin{figure}[t!]
	\centering
	\includegraphics[width=\linewidth]{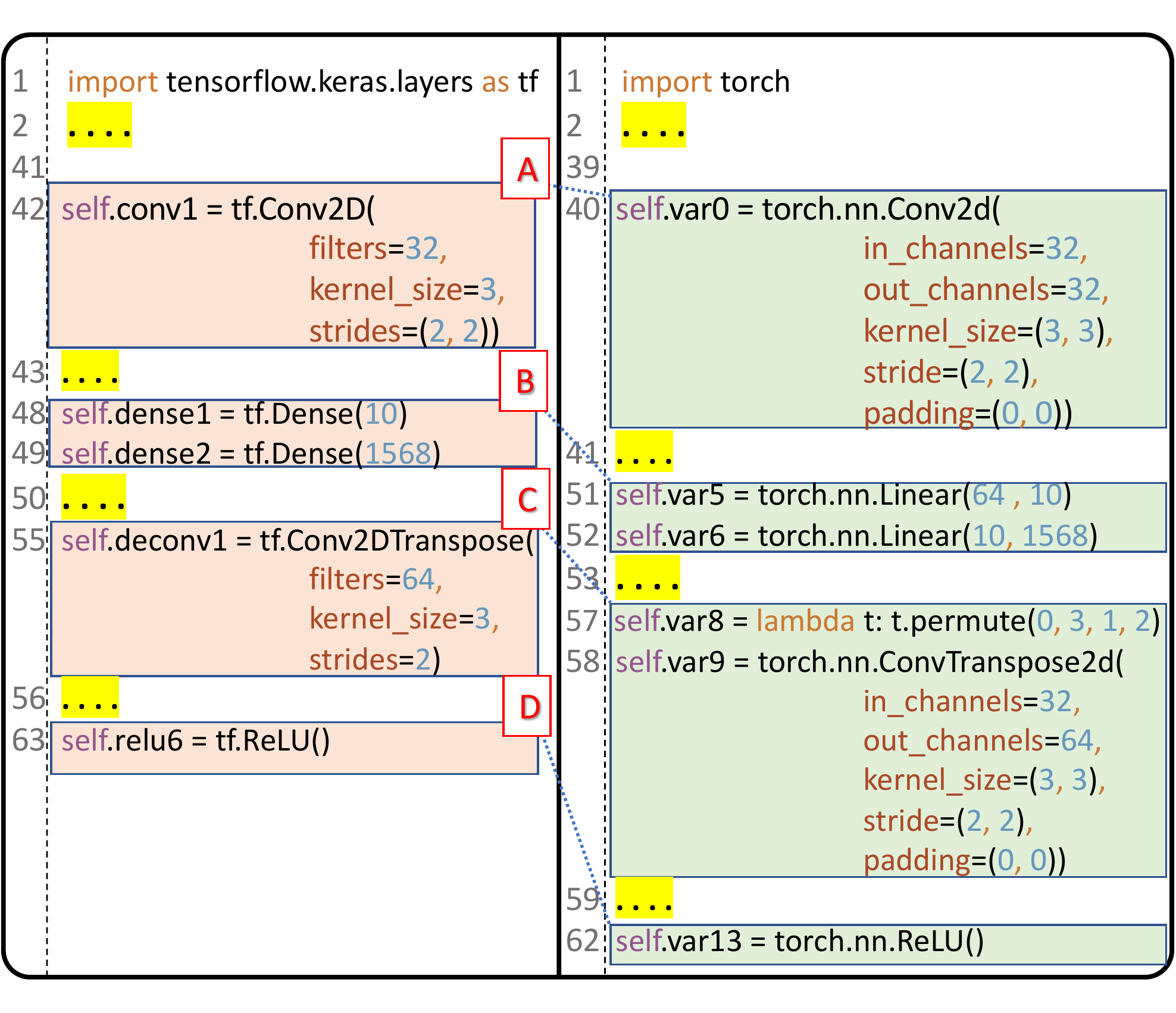}
	\caption{\small An example of how \method refactors a program written with \tf (left) to using \torch (right). Note that the whole program consists of 15 APIs calls to \tf, though we only show four blocks of them (\ie A, B, C and D) for brevity. \method can migrate the full program in 161 seconds. %\an{I think this figure should be on the first page?} \aidan{It should be close to where it's referred to first in the text, so 2nd page or 3rd page should be ok}}
	}
	\vspace{-3mm}
	\label{fig:example}
\end{figure}

We illustrate some of the difficulties of manual API refactoring via example.
Consider the TensorFlow code snippet on the left-hand-side of
Figure~\ref{fig:example}. The program being refactored shows an
autoencoder program~\cite{autoencoder} 
written using the TensorFlow API; the goal is to migrate this code to use the
PyTorch API.  
An autoencoder is a type of neural network that is trained to copy its input to
its output. Specifically in this example, the autoencoder tries to compress an image with an encoder and then the decoder will try to restore the original image. 

The example in Figure~\ref{fig:example} shows only a portion of the 
program, for didactic purposes. 
To build the first layer of the encoder, function \lt{Conv2D} is called, which constructs a convolution layer that can be applied to 2D images.
After further (elided) activation and convolution layers,
it calls \lt{Dense} to output a latent representation of the input image.
Decoding this output follows roughly the same procedure as the encoding, but
using \lt{Conv2DTranspose} instead of \lt{Conv2D}.  
The function \lt{ReLu} appears in both the encoder (not shown) and decoder, initializes a type of activation layer to ensure non-linearity of the neural network. 

The example of deep learning library code and translation in
Figure~\ref{fig:example} illustrates several of the core challenges in
refactoring open-source APIs, as well as opportunities to inform an automated
approach. First, the names of function calls implementing similar functionality
may be very similar or even identical (such as those in blocks A, C, and D), or
completely different (\eg \lt{Dense} versus \lt{Linear} in block B). 
If a
developer were performing this migration manually, they might reference the API
documentation.
For example, the
TensorFlow documentation describes the \lt{Conv2D} class as a ``2D convolution
layer (\eg spatial convolution over images)''~\cite{tensorflow}; the
corresponding PyTorch documentation for the \lt{Conv2d} call describes it
similarly, as a ``2D convolution over an input signal composed of several input
planes''~\cite{conv2d-doc}.  Here, the function names map well, but when this does 
not happen, it is more challenging to connect the documentation.
%and connecting the
%documentation requires somewhat deeper thought.

Even when we know which function to use, however, calls that
implement the same functionality can require different types, parameters, parameter
names, and even the parameter values may be different between them. 
This is true for
the majority of the calls in our example (see those in blocks A, B, and C).
Note for example that the \lt{Conv2D} functions take different parameters in
each of the two libraries.   There is overlap between
them --- both include \lt{kernel_size}, and \lt{stride} and \lt{strides} clearly
correspond --- but even the in-common parameters are not in the same argument
position between the two calls (\lt{strides} is the third parameter in
TensorFlow but \lt{stride} is the fourth in PyTorch).
Sometimes, some or all of the arguments to a call in the source API can be
copied directly to the call in the target API (see the calls in blocks A, C);
other times, correct arguments must be inferred (such as the first parameter to
\lt{Linear} in block B). 
Finally, in other situations, no single function in the target API
can match the semantics of a call from the source API, requiring instead a
one-to-many mapping (as we see in converting the \lt{Conv2DTranspose} call in
block C).

In the next section, we show how \method addresses these challenges with each of
its components.

\section{Refactoring algorithm}
\label{sec:Algorithm}
\begin{figure*}[t!]
    \centering
    \includegraphics[width=\textwidth]{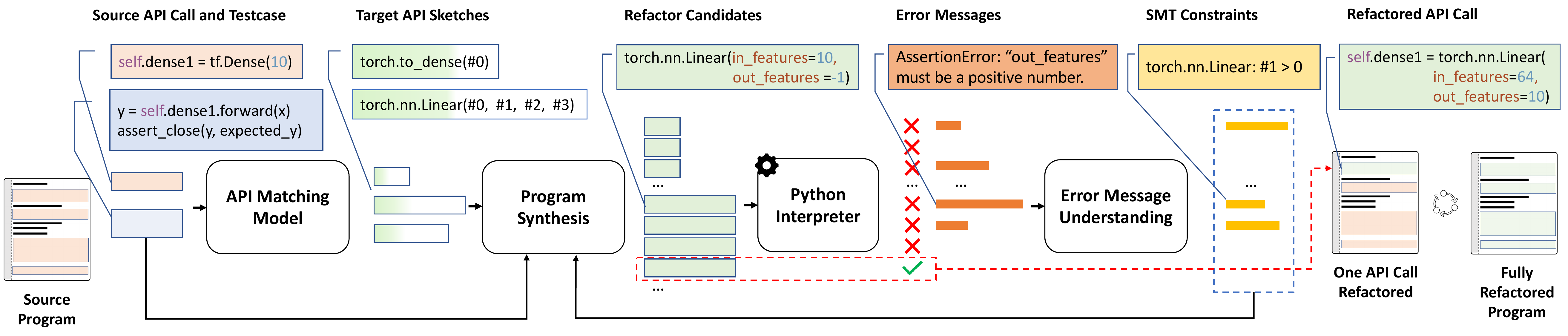}
    \caption{\small Overview of \method's architecture.}
    \label{fig:soar_overview}
    %\vspace{-5mm}
\end{figure*}

This section describes \method, our approach for automatic API
migration. We begin with a high-level overview of the method
(Section~\ref{sec:overview}) before providing more detail on individual
components (Section~\ref{sec:matching}; ~\ref{sec:synthesis}; ~\ref{sec:errors}).
%, and conclude with
%a few comments on our implementation (Section~\ref{sec:implementation}).  

\begin{algorithm}[!t] 
\caption{\textsc{Synthesizer}($\mathcal I, \mathcal S, \mathcal T, \mathcal C$)}
\label{alg:synthesizer}
\begin{algorithmic}[1]
\Require $\mathcal I$: existing program, 
$\mathcal S$: source library, 
$\mathcal T$: target library, $\mathcal C$: test cases
\Ensure $\mathcal O$: refactored program
\State $\vec{r}$ : API mapping = \textsc{mapAPI}($\mathcal{T}, \mathcal{S}$)
\State $\mathcal O = [~]$
\ForEach {$l \in \mathcal I $}
\State $\mathcal O = \mathcal O + [\textsc{refactorLine}(l, \mathcal T, \mathcal C, \vec{r})]$
\EndFor
\end{algorithmic}
\end{algorithm}

\subsection{Overview}
\label{sec:overview}

\begin{figure}
	\centering
	\includegraphics[width=0.48\textwidth]{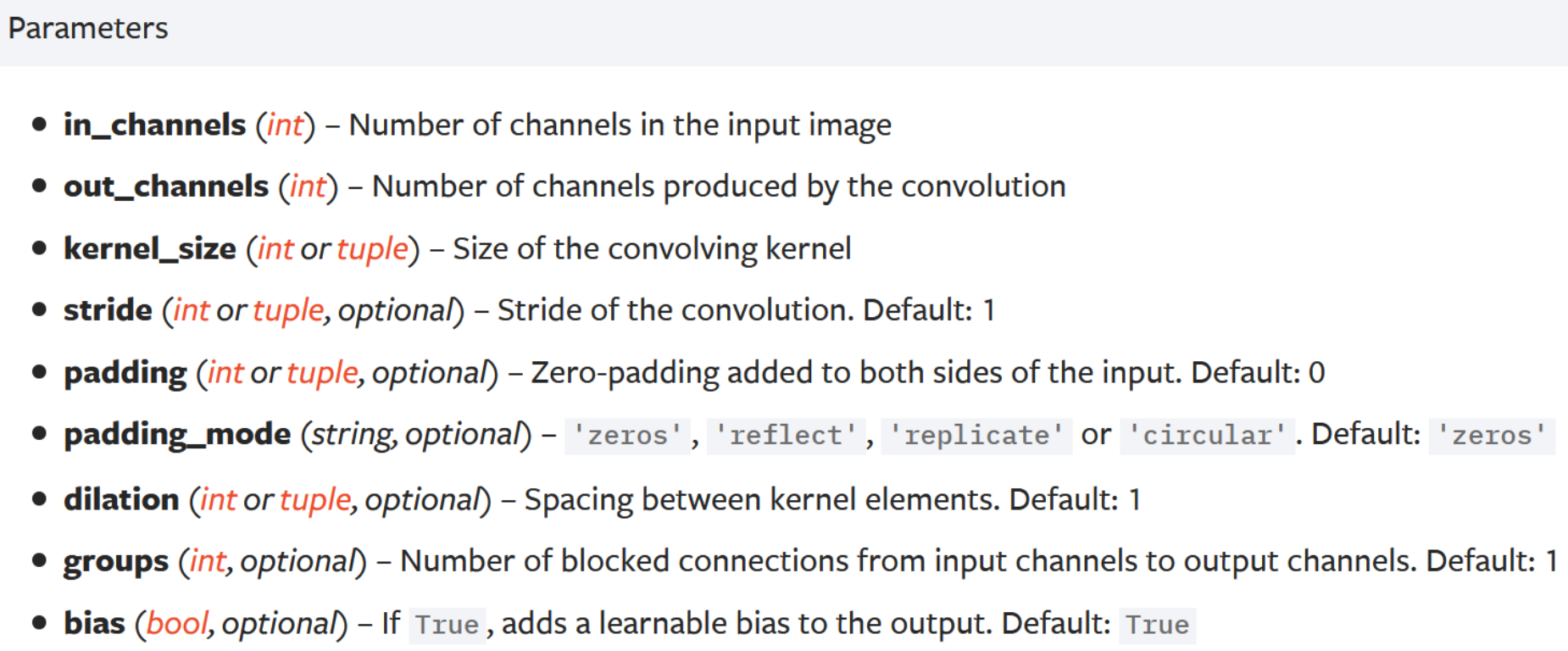} 
	\caption{\small Description of the program parameters in \lt{torch.nn.Conv2d} documentation \cite{conv2d-doc}.}
	\label{conv2d_param}
	%\vspace{-2mm}
\end{figure}

Figure~\ref{fig:soar_overview} shows an overview of the \method architecture,
while Algorithm~\ref{alg:synthesizer} provides an algorithmic view. \method
takes as input a program $\mathcal I$ consisting of a sequence of API calls from
a source library $\mathcal S$, the source ($\mathcal S$) and target ($\mathcal T$) libraries and their corresponding documentation, and a set of existing test cases ($\mathcal C$). Since the user wants to refactor code from $\mathcal S$ to $\mathcal T$, we assume that the user already has test cases for $\mathcal I$ that can be reused to check if the refactored code ($\mathcal O$) has the same functional behavior has the original code ($\mathcal I$). 
%
%\clg{I think a sentence or two somewhere near here
%  explaining why it's completely reasonable to assume test cases
%  would be good, to preempt reader crankiness.  I'm struggling
%  with phrasing, though, hence this todo.} \an{I am not sure how to address this, I thought this is standard with synthesis? Maybe Ruben has better intuition on this?} 
%The overall goal of the approach is to produce a refactored
%program $\mathcal O$ that has the same functionality as $\mathcal I$, and passes
%the same tests, but instead uses library $\mathcal T$.  
%
Refactoring proceeds one line at a time in $\mathcal I$, finding/constructing an
equivalent snippet of code (composed by one or more lines) that uses APIs of the
target library $\mathcal T$; the composition of all these translated lines
comprises the output $\mathcal O$.

For each API call in the input program, the first problem either a developer or
a tool must face is to identify methods in the target API that implement the
same functionality (\ie for a given set of input parameters, the target API
call must generate the same output). \method uses an \textit{API matching model}
to identify target API calls. This
model is built using NLP techniques that analyze the provided API documentation
for each call, and provides a mapping ($\vec{r}$ in
Algorithm~\ref{alg:synthesizer}) that computes the similarity between each
target API function and each potential source API function. \method uses this to
find the most likely replacement methods in the target API for each source API
call in the input program. We provide additional detail in
Section~\ref{sec:matching}.

\looseness-1
Given a potential match call in the target API, the next step is to determine
\emph{how} to call it, in terms of providing the correct parameters, in the
correct order, of the correct type.  
\method uses program synthesis to automatically write the refactored
API call, using the provided test cases to define the expected behavior of
the synthesized code and its constituent parts.  The synthesis process
can be assisted with additional automated analysis of API documentation, which often
provides key information about each parameter, namely (1) whether it is required
or optional, (2) its type, (3) its default value (if applicable), and (4) constraints between arguments, input and output (\eg input and output tensor shapes). 
%\an{added (4), but not sure whether its the right place to add it} \aidan{I don't think my Fig3 example can convey point 4, so maybe this will confuse reader?}
%
Figure~\ref{conv2d_param}
shows a snippet of the descriptions of all parameters for \lt{torch.nn.Conv2d}.  For example, the parameter \lt{stride} is optional; it takes type
\lt{int} or \lt{tuple}, and its default value is 1.  
Analysis of this documentation can produce a \emph{specification constraint} for
the \lt{stride} parameter, assisting the program synthesis task. 
Section~\ref{sec:synthesis} describes the synthesis step.

Given a potential rewrite in the target API, a natural step for a developer
would be to run the refactored code on test inputs.  
Unsuccessful runs can be quite informative, because many APIs (especially in the
deep learning and data science domains) provide error messages that can be very
helpful for debugging.  \method simulates the manual debugging process by
first adapting the input whole-program test cases to test partially refactored
code, and then extracting both syntactic and semantic
information from any error messages observed when running them.  \method uses
this information to add new constraints to the
iterative synthesis process (Section~\ref{sec:errors}).
%; we describe how in Section~\ref{sec:errors}.   

After migrating all calls in the source API to the target API such that all
input tests pass, \method outputs a fully refactored program. Subsequent
sections provide additional detail on the previously described steps. 

\subsection{API Representation Learning and Matching}
\label{sec:matching}

The first step in migrating a call in a source API is to identify candidate
replacement calls in the target API with similar semantics.  The \emph{API
  matching model} supports this task by analyzing the prose documentation
associated with each call in each API, and computing similarity scores between
all API pairs.
At a high level, this model embeds each API method call in a source and target
library into the same continuous high-dimensional space, and then computes
similarity between two calls in terms of the distance between them in that
space. We explored two ways on obtaining API representation: TF-IDF (term frequency -- inverse document
frequency)~\cite{salton1988term} and pretrained word embeddings~\cite{pennington2014glove}. 
%\an{So I falled back and restore the structure a bit. I think we still need to include the names of the representation methods because we directly compare them in the evaluations. I do find myself including too much detail and can definitely trim it down if necessary.}

\vspace{1ex}
\noindent\textbf{TF-IDF.}
The intuition behind TF-IDF is to find the most \emph{representative} words rather than the most frequent words in a sentence.
Normalizing by the inverse-document-frequency 
lowers the weights of common keywords that are less informative, such as
\textit{torch}, \textit{tensorflow} and those stop words in natural language
such as \textit{the} or \textit{this}.

Specifically, we first derive a bag-of-words representation $\mathbf{x^i}$ from a description of an API call after some stemming of the words with the Snowball Stemmer \cite{porter2001snowball}.
$\mathbf{x^i} = [x_1^i, x_2^i, ..., x_n^i]$ where $x_j^i$ denotes the frequency
with which word $x_j$ appeared in the sentence $\mathbf{x^i}$, and $n$ is the size of the
vocabulary from the descriptions of all APIs we are trying to embed. A
TF-IDF representation of the call is computed as Equation \ref{tf-idf}:

\vspace{-2mm}
\begin{equation}
\label{tf-idf}
    \text{TF-IDF}(\mathbf{x^i}) = 
    \left[ \frac{x_1^i}{\sum_{t=0}^{m} x_1^t}, 
     \frac{x_2^i}{\sum_{t=0}^{m} x_2^t}, ..., 
     \frac{x_n^i}{\sum_{t=0}^{m} x_n^t} \right]
\end{equation}

However, the major downside of TF-IDF is that it does not encode the similarities between words
themselves. For example, consider two hypothetical call descriptions:
(1) \textit{Remove the last item of the collection}, and (2) \textit{Delete one element from the end of the list}. 
They are semantically similar but since they have minimal overlapping words, a TF-IDF
representation method would not recognize these two API calls as similar.

\vspace{1ex}
\noindent\textbf{Tfidf-GloVe.}
\reb{We can extend the TF-IDF representation to recognize similar words by adding pretrained word embeddings}. Specifically, we propose to use the GloVe embedding~\cite{pennington2014glove}, which
is trained on a very large natural language corpus and learns to embed similar words
closer in the embedding space.

To obtain sentence embeddings from individual words, we perform a weighted average of the word embeddings and use the TF-IDF scores of individual words as weight factors. It is a simple yet effective method to obtain sentence embedding for downstream tasks, as noted by previous work ~\cite{perone2018evaluation,arora2016simple}.
This is shown in detail as Equation \ref{embedding}, where
$\mathbf{w_j}$ is the vector encoding the GloVe embedding of word $x_j$: 

\vspace{-1mm}
\begin{equation}
\label{embedding}
    \text{Embedding}(\mathbf{x^i}) = 
    \sum_{i=j}^{n} \frac{x_j^i\cdot \mathbf{w_j}}{\sum_{t=0}^{m} x_j^t}
\end{equation}
 
By including the GloVe embedding, word similarity is preserved; by including the
TF-IDF terms, the influence of embeddings of common words is greatly reduced. However, GloVe is trained with Common Crawl~\cite{common-crawl} which contains raw webpages, which is a mismatch from our domain of textual data (\ie data science and programming). %This causes a lot of OOV (out-of-vocabulary) problems.

\vspace{1ex}
\noindent\textbf{API matching.}
Given the representation of two APIs $\text{Rep}(\mathbf{x}^i) $, $\text{Rep}(\mathbf{x}^j)$ in the same space $\text{Rep}(\cdot)$, we compute their similarity with cosine distance:
\begin{equation}
    \text{sim}(\text{Rep}(\mathbf{x}^i), \text{Rep}(\mathbf{x}^j)) = \frac{\text{Rep}(\mathbf{x}^i) \cdot \text{Rep}(\mathbf{x}^j)}{|\text{Rep}(\mathbf{x}^i)||\text{Rep}(\mathbf{x}^j)|}
\end{equation}

For computational efficiency, we pre-compute the similarity matrix between the APIs across the source and target library. So we will be able to query the most similar API for the synthesizer to synthesize its parameters on the fly.

\subsection{Program Synthesis}
\label{sec:synthesis}

Given the
input test cases and the API matching model providing 
a ranked list $\vec{r}$ of APIs in the target library, the synthesis model
automatically constructs new, equivalent code, of one or more lines, that uses
APIs of the target library $\mathcal T$. The  
refactored program $\mathcal O$ has the same functionality as input
program $\mathcal I$, and passes the same set of tests
$\mathcal C$.

To refactor each line of the existing program $\mathcal I$, we use techniques 
of programming-by-example (PBE) synthesis~\cite{DBLP:journals/ftpl/GulwaniPS17}. 
PBE is a common 
approach for program synthesis, where the synthesizer takes as specification a 
set of input-output examples and automatically finds a program that satisfies 
those examples. In the context of program refactoring, our examples
correspond to the test cases for the existing code.
In this paper, we restrict ourselves to 
straight-line code where each line returns an object that can be tested.
With these assumptions, we can automatically generate new test cases for each 
line $k$ of program $\mathcal I$. This can be done by using the input of the 
existing tests, running them, and using the output of line $k$ as a new test 
case for the program composed by lines $1$ to $k$. 

\begin{algorithm}[!t] 
\caption{\textsc{refactorLine}($l, \mathcal T, \mathcal C, \vec{r}$)}
\label{alg:refactorline}
\begin{algorithmic}[1]
\Require $l$: line of code from $\mathcal I$, 
$\mathcal T$: target library, $\mathcal C$: test cases,
$\vec{r}$: ranked list of API matchings
\Ensure $\mathcal R$: refactored snippet
\ForEach {$a \in {\vec{r}[l]} $} \algorithmiccomment{$a$ is a target API}
\State $\vec{s}$ = \textsc{generateSketches}($a, \mathcal T$)\label{li:sketch}
\ForEach {$s \in \vec{s}$}
\State $\mathcal R$ = \textsc{fillSketch}($s$)\label{li:fill}
\If {\textsc{passTests}($\mathcal R, \mathcal C$)}\label{li:tests}
\State \Return $\mathcal R$
\EndIf
\EndFor
\EndFor
\end{algorithmic}
\end{algorithm}

Our program synthesizer for refactoring of APIs is presented in 
Algorithm~\ref{alg:refactorline} and it is based on two 
main ideas: (i) program sketching, and (ii) program enumeration. For each line 
$l$ in program $\mathcal I$, we start by enumerating a program sketch (\ie 
program with holes) using APIs from the target library $\mathcal T$ 
(line~\ref{li:sketch}). For each program sketch, we perform program 
enumeration on the possible completion of the API parameters (line~\ref{li:fill}). For each complete program, we run the test cases for the 
program up to line $l$. If all test cases succeed, then we found a correct 
mapping for line $l$ between libraries $\mathcal S$ and $\mathcal T$ (line~\ref{li:tests}). 
Otherwise, we continue until we find a complete program that passes all test 
cases.

\vspace{1ex}
\noindent \textbf{Program Sketching.} Program sketching is a well-known technique 
for program synthesis \cite{DBLP:conf/aplas/Solar-Lezama09} where the programmer provides a sketch of 
a program and the program synthesizer automatically fills the holes in this 
sketch such that it satisfies a given specification. We refactor \emph{one line
} of program $\mathcal I$ at each time. Our first step is to use the ranked 
list of APIs to create a program sketch where the parameters are unknown.
For instance, consider the first layer from the motivating example that shows the network for an autoencoder using TensorFlow:

\noindent \lt{tf.keras.layers.Conv2D}\\
\lt{(filters=32, kernel_size=3,strides=(2, 2))}

A possible sketch for this call using PyTorch is:

\noindent \lt{torch.nn.Conv2d}\\
\lt{(#1,#2,(#3,#4),stride=(#5,#6),padding=(#7,#8))}

\looseness-1
Where holes {\footnotesize \texttt{\#i}} have to be filled with a specific value for the APIs 
to be equivalent.
This approach works for \emph{one-to-one} mappings but would not support common \emph{one-to-many} mappings where the parameters often need to be transformed before being used 
in the new API. This is the case of the previous API where a reshaping operation must be performed before calling the PyTorch API. 
To support this common behavior, we include in our program 
sketch \emph{one} API from the target library $\mathcal T$ and common 
reshaping APIs (\eg permute, long). 

The sketch that corresponds to the refactoring solution of the \lt{Conv2D} API from TensorFlow uses a reshaping API before calling the \lt{Conv2d} API from PyTorch:

\noindent \lt{lambda x: x.permute(#9,#10,#11,#12)}\\
\lt{torch.nn.Conv2d}\\
\lt{(#1,#2,(#3,#4), stride=(#5,#6),padding=(#7,#8))}

Using Occam's razor principle, our program synthesizer enumerates program 
sketches of size $1$ and iteratively increases the size of the synthesized 
program up to a specified limit. 

\vspace{1ex}
\noindent \textbf{Program Enumeration.} For each program sketch $\mathcal P$, 
our program synthesizer enumerates all possible completions for each hole. 
Since each hole has a given type, we only want to enumerate well-typed 
programs. We encode the enumeration of well-typed programs into a 
Satisfiability Modulo Theories (SMT) problem using a combination of Boolean 
logic and Linear Integer Arithmetic (LIA). 
This encoding is similar to other approaches that use SMT-based enumeration 
for program synthesis \cite{DBLP:conf/cp/OrvalhoTVMM19, DBLP:journals/pvldb/MartinsCCFD19} and encodes the following properties:

\begin{itemize}
	\item Each hole contains exactly one parameter;
	\item Each hole only contains parameters of the correct type.
\end{itemize}

A satisfying assignment to the SMT formula can be translated into a complete 
program. The types for each hole can be determined by extracting this 
information from documentation, by performing static analysis, or by having 
this information manually annotated in the APIs. The available parameters and 
their respective types can be extracted automatically from the parameters used 
in the $k$-th line of program $\mathcal I$ and by any default parameters that 
can be used in the API from $\mathcal T$ that appears in the program sketch 
$\mathcal P$. For instance, for the \lt{Conv2d} example presented in this section, we consider as possible values for the holes, the values that appear in the existing code (32, 3, 2) and default values for integer parameters (-1, 0, 1, 2, 3) that are automatically extracted from documentation.

%the \texttt{permute} and \texttt{Conv2d} APIs: (1,2,3,4)

Encoding the enumeration of well-typed programs in SMT has the advantage of 
making it easier to add additional logical constraints that can prune the search space.

\vspace{1ex}
\noindent \textbf{Specification Constraints.}
As we described in Section~\ref{sec:overview}, API documentation often provides
additional useful information about parameters to function calls, including type
and default values.  For each considered API call, we scrape/process the
associated documentation to extract these properties and encode them as SMT
constraints to further limit the synthesizer search space. 

\begin{figure}[t]
	\centering
	\includegraphics[width=.48\textwidth]{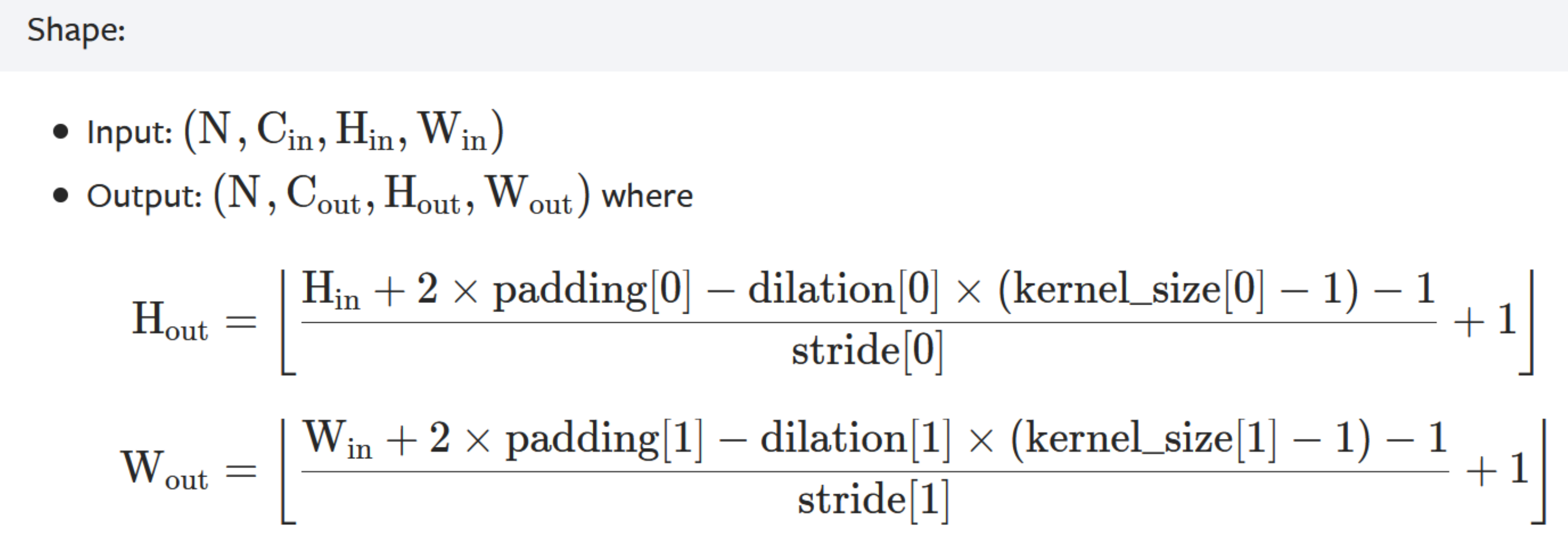} 
	\caption{\small Relationship between the parameters of \texttt{Conv2d} API described in PyTorch documentation~\cite{conv2d-doc}.}
	\label{fig:conv2d:relationship}
	\vspace{-3mm}
\end{figure}

Additionally, some APIs have complex 
relationships between parameters which if encoded into SMT may reduce the 
search space considerably. For instance, Figure~\ref{fig:conv2d:relationship} shows the relationship between the different parameters for the \lt{Conv2d} API described in PyTorch documentation.
For APIs with these kinds of shape constraints, we can 
encode these relationships into SMT to further prune the 
number of feasible completions. When we use these relationships in our
experiments, we encode them manually (a one-time cost for an actual \method user
or API maintainer), but we observe that in many cases they could be
automatically extracted from documentation.

Besides these specification constraints, we can also further prune the search 
space by using the error messages provided by the Python 
interpreter, as we discuss in the next section.

% \textbf{Daniel's section}

% We synthesize a target program for all candidate API calls. We encode the enumeration of well-typed programs into SMT and use Z3 to enumerate well-typed programs \cite{de2008z3}. We choose the target program with the same output as the source program (e.g., the target program has the same dimension output tensor as the source program). If the target program fails during run-time, we attempt to generate SMT constraints to limit the synthesizer search space.

\subsection{Error Message Understanding}
\label{sec:errors}

\begin{figure}[t]
	\centering
	\includegraphics[width=.45\textwidth]{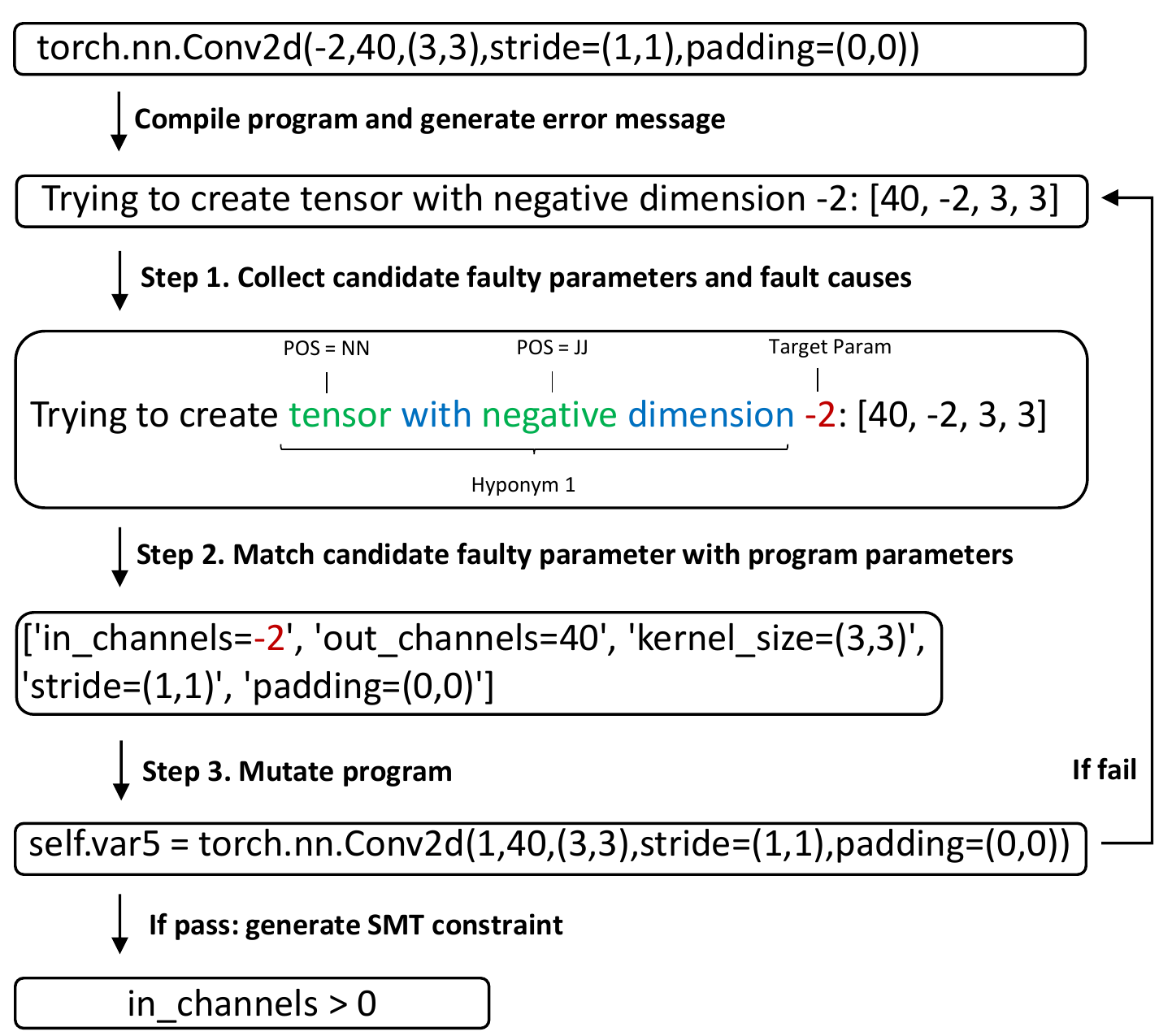} 
	\caption{\small Example error message to SMT constraint pipeline using hyponym 1.}
	\label{z3_gen}
	\vspace{-3mm}
\end{figure}

We use a combination of extracting hyponymy relations and Word2vec~\cite{DBLP:conf/nips/MikolovSCCD13} to understand run-time error messages.
As outlined in Figure \ref{z3_gen}, our SMT constraint generation method consists of three steps.

\begin{table*}[t]
  \centering
  \caption{The four hyponyms in the error message understanding model}
  \label{tab:hyponyms}
  \begin{tabular}{p{1cm}lp{4.6cm}p{3.5cm}}
    \toprule
    Type & NP & Example error messages & Identified hyponym \\
    \midrule
    1 & $\{\langle Noun \rangle * \langle Preposition  \rangle \langle Adjective \rangle ? \langle Noun \rangle\}$ & `Trying to create tensor with negative dimension -1: [-1, 100, -1, -1]' & tensor with negative dimension \\
    2 & $\{\langle Noun \rangle \langle Cardinal\_number \rangle\}$ & `embedding(): argument weight (position 1) must be Tensor, not int' & position 1 \\
    3 & $\{\langle Coordinating\_conjunction \rangle \langle Verb \rangle  \langle Adjective \rangle \langle Noun \rangle\}$ & `Expected 3-dimensional input for 3-dimensional weight [2, 2, 3],
    but got 4-dimensional input of size [100, 50, 40, 1] instead' & but got 4-dimensional input \\
    4 & $\{\langle Verb \rangle \langle Adverb\rangle \langle Verb\_past\_participle \rangle\}$ & `non-positive stride is not supported' & is not supported \\
    
    \bottomrule
  \end{tabular}%
  \vspace{-2mm}
\end{table*}{}

\vspace{1ex}
\noindent\textbf{Step 1: Extract hyponymy relation candidates from error messages.}
We perform an automatic extraction of customized hyponyms on each error message.  Hyponyms are specific lexical relations that are expressed in well-known ways \cite{hearst1992automatic}. In encoding a set of lexico-syntactic patterns that are easily recognizable (\ie hyponyms), we avoid the necessity for semantic extraction of a wide-range of error message text. We then use the collected hyponyms to map the error message to a single faulty parameter, and output a SMT constraint based on the faulty parameter. 

Prior work on text parsing uses Tregex, which is a utility developed by Levy and Andrew for matching patterns in constituent trees \cite{levy2006tregex}. For example, Evans \textit{et al.} evaluated the performance of Tregex on privacy policies \cite{evans2017evaluation}. However, Deep Learning (DL) API compilation error messages are domain specific. Sumida \textit{et al.} used the hierarchical layout of Wikipedia articles to  identify hyponymy relations \cite{sumida2008hacking}. Similarly to Wikipedia documents, DL API compilation error messages are more consistent and organized than normal, natural language, documents. Therefore, we follow the approach of extracting hyponymy relations based on the hierarchical layout of a string.

We propose a set of four lexico-syntatic patterns to identify hyponyms using
\textit{noun-phrases} (NP) and regular expressions frequently appearing in
machine learning API error messages. Table \ref{tab:hyponyms} shows the four hyponyms.
If we identify any of the four lexico-syntatic patterns within an error message, we tag the error message with a hyponym type. As shown in Figure \ref{z3_gen}, we identify hyponym 1 in error message ``\textit{Trying to create tensor with negative dimension...}". 
%  Noun-phrases are the results of \textit{chunking}, which uses part-of-speech (POS) tags as input and outputs \textit{chunks} of POS tags \cite{pvs2007part}. We discover the four lexico-syntatic patterns from the observation of  POS tagging on the API mapping error messages.
% Below is the list of lexico-syntatic patterns that indicate the domain specific hyponymy relations for the error message understanding pipeline.

\vspace{1ex}
\noindent
\textbf{Step 2: Identify candidate faulty parameters and constraints.}
Step 2 uses different keywords based on the result of step 1 to identify the faulty parameter. As shown in Figure \ref{z3_gen}, an error message with hyponym 1 is likely to have the POS=JJ word as a parameter constraint (\ie word ``\textit{negative}"). Based on the fault cause candidate, we then store all negative numbers as candidate faulty parameters (\eg [40, -2, 3, 3] has -2 as the only faulty parameter).
We then vectorize the candidate faulty parameter name (\ie -2) and find the program parameter name with the closest vectorized distance. As shown in Figure \ref{z3_gen}, the parameter ``$in\_channels=-2$" has the nearest vectorized distance to the candidate faulty parameter -2. Based on the fault cause, we generate a candidate constraint. The example error message in Figure \ref{z3_gen} has only one candidate constraint: ``$in\_channels >= 0$".   

\vspace{1ex}
\noindent
\textbf{Step 3: Mutate program.}
To validate the candidate faulty parameters and constraints, we mutate each faulty parameter according to each faulty parameter and constraints pair. We then re-compile the program for each mutation. If the error message remains the same, we discard the faulty parameter and constraint pair as a candidate. If the program passes, or if the error message changes, we store the faulty parameter and constraint pair as an SMT constraint.
As shown in Figure \ref{z3_gen}, the API call mutator mutates the second parameter (``$in\_channels=-2$") to a non-negative number. The mutator first attempts ``$in\_channels = 0$" and it encounters a different error message. From the new error message, we mutate this parameter to ``$in\_channels = 1$" and observe no further errors. Therefore, we refine our previous constraint to be ``$in\_channels > 0$'', and store it as the final SMT constraint for the program in Figure \ref{z3_gen}.

\section{Evaluation}
\label{sec:Evaluation}
\reb{We selected two migrations tasks: \tf to \torch and \dplyr to \pandas. 
We believe that these two migration tasks are representative of the needs of the data-science community. Indeed, \tf and \torch are the two most popular deep learning frameworks, and recent trends indicate that a large portion of \tf user-base is shifting to \torch \cite{he2019mlframeworks}. We thus chose it as an indicative, relevant task. Similarly, \dplyr is one of the top-5 most downloaded {R} libraries; \pandas is its {python} counterpart. } 

To evaluate our approach we answer the following research questions:

\begin{enumerate}[label=\textbf{{Q}\arabic{*}.}]
	
	\item How effective is \method at migrating neural network programs between different libraries?
	\item How does each component of \method impact its perfomance?
	\item Is \method generalizable to domains besides deep learning library migration?

\end{enumerate}

\subsection{Benchmarks and experimental setup}

We collected $20$ benchmarks for each of the two migration tasks. In particular,
for the \tf to \torch task, we gathered 20 neural network programs from tensorflow tutorials~\cite{tenstorflow-tutorial}, off-the-shelf models implemented with \tf~\cite{tenstorflow-apps} or its model zoo~\cite{tenstorflow-models}.
This set of benchmarks includes: Autoencoders for image and textual data, classic feed-forward image classification networks (\ie the VGG family, AlexNet, LeNet, etc), convolutional network for text, among others. The average number of layers in our benchmark set is $11.80 \pm 11.52$, whereas the median is $8$. Our largest benchmark is the VGG19 network which contains $44$ layers.

For the domain of table transformations, we collected $20$ benchmarks from Kaggle~\cite{kaggle}, a popular website
for data science. The programs in the benchmark set have an average of $3.05 \pm 1.07$ lines of code, and a median of $3$ lines. Although the programs considered for this task are relatively small compared to the deep learning benchmarks, they are still relevant for data wrangling tasks as shown by previous program synthesis approaches \cite{DBLP:conf/pldi/FengMGDC17}.

\reb{Each benchmark is also associated with a set of input-output examples (i.e., test cases) used to decide migration success. For the deep learning task, the test cases are automatically generated by running the original neural network on random inputs. Whereas the test cases for the \dplyr to \pandas task are user provided.}

All results presented in this section were obtained using an Intel(R) Xeon(R) 
CPU E5-2630 v2 @ 2.60GHz, with 64GB of RAM, running Debian GNU/Linux 10, and a time limit of 3600 seconds. To evaluate the impact of each component in \method, we run four versions of the tool. \method with TF-IDF (\method w/ TF-IDF) and \method with tfidf-GloVe (\method w/ Tfidf-GloVe) to evaluate the impact of API representation learning methods. \method without specification constraints (\method w/o Specs.) and \method without error message understanding (\method w/o Err. Msg.) to evaluate the impact of these components on the performance of \method.

\subsection{Implementation}
The \method implementation integrates several technologies.
Scrapy \cite{scrapy}, a Python web-scraping framework, is used to collect documentation for the four libraries in our experiments. 
To enumerate programs in the synthesis step, we use the Z3 SMT solver \cite{DBLP:conf/tacas/MouraB08}. For each target program call parameter, we extract an answer for the four parameter questions in Section \ref{sec:overview} and generate corresponding SMT constraints. 
In both API matching model and the error message understanding model, the GloVe word embeddings \cite{pennington2014glove} are used as an off-the-shelf representation of words. 
For the four libraries appearing in our two evaluation migration tasks, we use \tf 2.0.0, \torch 1.4.0, \dplyr 1.0.1 (with R 4.0.0) and \pandas 1.0.1, though our proposed method and associated implementation do not rely on specific versions. \reb{We provide a replication package, including benchmarks, source code and virtual environment to run SOAR.\footnote{https://zenodo.org/record/4452730}}

\begin{table}[t]
	\centering
	\caption{Execution time for the deep learning library migration task in each of the $20$ benchmarks. }
	\label{tab:dl_spec_err_msg}
    \resizebox{\linewidth}{!}{%
    \begin{tabular}{lrrr}
    \toprule
                                               & SOAR            & \method w/o Specs. & \method w/o Err. Msg. \\ \midrule
    conv\_pool\_softmax(4L)                    & \textbf{1.60}   & 23.02                  & 14.35                        \\
    img\_classifier(8L)       & \textbf{12.82}  & 336.00                 & 65.66                        \\
    three\_linear(3L)                          & 3.18            & \textbf{2.34}          & 21.07                        \\
    embed\_conv1d\_linear(5L)                  & \textbf{5.27}   & 123.85                 & 16.90                        \\
    word\_autoencoder(3L)                      & 1.81            & \textbf{1.46}          & 2.64                         \\
    gan\_discriminator(8L)                     & \textbf{12.80}  & timeout                & 252.20                       \\
    two\_conv(4L)                              & 16.69           & timeout                & \textbf{15.09}               \\
    img\_autoencoder(11L)                      & \textbf{160.97} & 391.09                 & 487.54                       \\
    alexnet(20L)                               & 425.22          & timeout                & \textbf{66.13}               \\
    gan\_generator(9L)                         & \textbf{412.47} & timeout                & timeout                      \\
    lenet(13L)                                 & \textbf{280.91} & timeout                & timeout                      \\
    tutorial(10L)                              & \textbf{6.04}   & timeout                & 58.29                        \\
    conv\_for\_text(11L) & \textbf{9.04}   & timeout                & 32.29                        \\
    vgg11(28L)                                 & \textbf{40.83}  & timeout                & 132.67                       \\
    vgg16(38L)                                 & \textbf{82.05}  & timeout                & 139.27                       \\
    vgg19(44L)                                 & \textbf{83.99}  & timeout                & 189.90                       \\
    densenet\_main1(5L)                        & timeout         & timeout                & timeout                      \\
    densenet\_main2(3L)                        & timeout         & timeout                & timeout                      \\
    densenet\_conv\_block(6L)                  & timeout         & timeout                & timeout                      \\
    densenet\_trans\_block(3L)                 & timeout         & timeout                & timeout  \\\bottomrule
    \end{tabular}}
    \vspace{-2mm}
\end{table}{}

\subsection{Q1: \method effectiveness} 

Table \ref{tab:dl_spec_err_msg} shows how long it takes to migrate
each of the deep learning models from \tf to \torch, using the various approaches. Our best approach (shown as \method) successfully migrates $16$ of the $20$ DL models with a mean run-time 
of $97.23 \pm 141.58$ seconds, and a median of $14.76$ seconds. 
The average number of lines in the $16$ benchmarks that we successfully migrate is $13.6 \pm 12.14$, %median 9.5
whereas the average number of lines in the output programs is $18.56 \pm 16.40$. %\an{are those numbers for the output (target) program? if so, we need to be clear about this}%median 12.5 
The reason the number of synthesized lines is higher than those in the original benchmarks is that we frequently do one-to-many mappings. In fact, $15$ out of the $16$ require at least one mapping that is one-to-many. In the $16$ benchmarks, \method tests on average $4414.18 \pm 5676$ refactor candidates (i.e. program fragments tested for each mapping), and it needs to test a median $2111$ candidates before migrating each benchmark. The reason $4$ benchmarks timeout is that in each of these benchmarks there is at least one API in the benchmark that has a poor ranking (\ie not in the top 200). 

\subsection{Q2: performance of each \method component}  

We perform an ablation study to understand the effectiveness of several features in the \method design.

\vspace{1ex}
\noindent\textbf{Embeddings.} \reb{In Table \ref{tab:dl_embedding}, we show the execution time and average ranking for the correct API matchings for each benchmark, using different API representation learning methods, namely TF-IDF and tfidf-GloVe, as described in Section \ref{sec:Algorithm}.} We can see that for these tasks of \tf to \torch migration, using TF-IDF-based API matching model works better than adding pretrained GloVe embeddings. We believe this is because similar APIs are often named with same words(\eg \lt{Conv2DTranspose} vs. \lt{ConvTranspose2d}) or even identical name (\eg the APIs of creating a Rectified Linear Unit are both named as \lt{ReLU(...)}), for \tf and \torch. Thus simple word matching method like TF-IDF is suffice for API matching purposes. However, things are different for the second task we consider (see Section~\ref{sec:results:dplyr} for more details).

Another interesting result worth noticing is that although the synthesis time differs for the two approaches, the average rankings are quite similar for most of the benchmarks.
The reason is that despite the average rankings of correct target APIs being similar, the incorrect APIs ranked by the model before the correct one is different, and the time it takes to rule out those incorrect APIs varies greatly, determined largely by the number of parameters required for that API.

\vspace{1ex}
\noindent\textbf{Error Message Understanding.} As shown in Table \ref{tab:dl_spec_err_msg}, \method performs significantly better when using the error message understanding model. We can observe that without this component, two of the benchmarks that \method could solve would timeout at the 1 hour mark. For the $14$ benchmarks it still manages to solve, the synthesis time increases on average $4.66\times$.

The number of performed evaluations also increase substantially for each benchmark. For the $16$ benchmarks that \method successfully migrates, we evaluate an average of $43319.63 \pm 61259.62$ refactor candidates without the error message understanding model. This corresponds to a $9.81\times$ increase in the number of necessary evaluations when compared to the full \method method. In summary, we can significantly reduce the search space by interpreting error messages. 

\vspace{1ex}
\noindent\textbf{Specifications Constraints.} In Table \ref{tab:dl_spec_err_msg}, we also show the impact of specification constraints that describe the relationship between different parameters of a given API (see Section~\ref{sec:synthesis} for details). Even though, we only have these complex specifications for the $7$ most common APIs, the impact on performance is significant. Without these specification we can only solve $6$ out of $20$ benchmarks. Relating the arguments of the APIs helps \method to significantly reduce the number of argument combinations that it needs to enumerate.

\begin{table}[t]
	\centering
	\caption{Execution time and average API ranking for each of the $20$ benchmarks using TF-IDF and GloVe models. }
	\label{tab:dl_embedding}
	\resizebox{\linewidth}{!}{%
	\begin{tabular}{lrrrr}
\toprule
\multirow{2}{*}{}                          & \multicolumn{2}{c}{SOAR w/ TF-IDF}                             & \multicolumn{2}{c}{SOAR w/ Tfidf-GloVe}                    \\ \cmidrule(lr){2-3} \cmidrule(lr){4-5}
                                           & \multicolumn{1}{c}{Time(s)} & \multicolumn{1}{c}{Avg. Ranking} & \multicolumn{1}{c}{Time(s)} & \multicolumn{1}{c}{Avg. Ranking} \\ \midrule
conv\_pool\_softmax(4L)                    & 1.60                        & 1.0                              & \textbf{1.56}               & 1.0                              \\
img\_classifier(8L)                        & \textbf{12.82}              & 2.8                              & 31.04                       & 2.8                              \\
three\_linear(3L)                          & 3.18                        & 8.0                              & 7.70                        & 8.0                              \\
embed\_conv1d\_linear(5L)                  & \textbf{5.27}               & 2.4                              & 7.75                        & 2.4                              \\
word\_autoencoder(3L)                      & 1.81                        & 1.0                              & \textbf{1.52}               & 1.0                              \\
gan\_discriminator(8L)                     & \textbf{12.80}              & 3.5                              & 37.01                       & 2.8                              \\
two\_conv(4L)                              & 16.69                       & 1.0                              & \textbf{13.75}              & 1.0                              \\
img\_autoencoder(11L)                      & \textbf{160.97}             & 1.9                              & 166.34                      & 1.9                              \\
alexnet(20L)                               & \textbf{425.22}             & 2.0                              & 428.42                      & 2.0                              \\
gan\_generator(9L)                         & \textbf{412.47}             & 2.1                              & 1892.86                     & 2.6                              \\
lenet(13L)                                 & \textbf{280.91}             & \textbf{4.3}                     & timeout                     & 89.1                             \\
tutorial(10L)                              & \textbf{6.04}               & 2.4                              & 21.31                       & 2.4                              \\
conv\_for\_text(11L)                       & \textbf{9.04}               & 2.3                              & 14.08                       & 2.3                              \\
vgg11(28L)                                 & \textbf{40.83}              & 1.8                              & 73.92                       & 1.8                              \\
vgg16(38L)                                 & \textbf{82.05}              & 1.6                              & 114.41                      & 1.6                              \\
vgg19(44L)                                 & \textbf{83.99}              & 1.5                              & 114.98                      & 1.5                              \\
densenet\_main1(5L)                        & timeout                     & \textbf{172.8}                   & timeout                     & 285.6                            \\
densenet\_main2(3L)                        & timeout                     & \textbf{10.0}                    & timeout                     & 285.5                            \\
densenet\_conv\_block(6L)                  & timeout                     & \textbf{293.3}                   & timeout                     & 634.0                            \\
densenet\_trans\_block(3L)                 & timeout                     & \textbf{291.0}                   & timeout                     & 662.7                            \\     

\bottomrule
\end{tabular}}
\vspace{-2mm}
\end{table}{}

\subsection{Q3: \method generalizability.}
\label{sec:results:dplyr}

\begin{figure}
    \centering
    \begin{tikzpicture}[scale=0.95]
        \begin{axis}[
            ybar, axis on top, height=4cm, width=7.8cm, bar width=0.6cm,
            ymax = 20,
            ymin = 0,
            enlarge x limits=0.25,
            ylabel style={align=center},
            symbolic x coords={1-10, 11-100, 101+},
            xtick=data,
            %x tick label style={font=\small,text width=1cm,align=center},
            ylabel=\small{\# benchmarks},
            xlabel=\small{Average ranking},
            legend style={at={(0.05,0.7)},anchor=west}
            ]
            \addplot[fill=blue!60!black!40, postaction={
        pattern=vertical lines}] coordinates {
                (1-10,0)
                (11-100,7)
                (101+,13)
            };
            \addlegendentry{TF-IDF};
            
            \addplot[fill=red!60!white!40, postaction={
        pattern=horizontal lines}] coordinates {
                (1-10,6)
            	(11-100,10)
            	(101+, 4)
            };
            \addlegendentry{Tfidf-GloVe};
        \end{axis}
    \end{tikzpicture}
	\caption{Average ranking of the APIs for each of the 20 dplyr-to-pandas benchmarks.}
	\label{fig:hist_dplyr}
	\vspace{-2mm}
\end{figure}
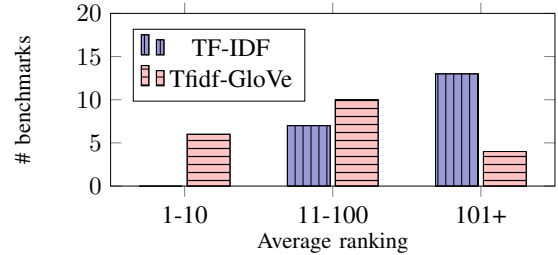

 Our experiments so far concern deep learning library migration in Python. 
To study the generality of our proposed method, we applied \method to another task of migrating from \dplyr, a data manipulation package for R, to \pandas, a Python library with similar functionality. Fig. \ref{fig:hist_dplyr} shows how the two API matching methods perform in this domain. While with Tfidf-GloVe, 30\% of the correct APIs are ranked among the top 5, saving lots of evaluations for the synthesizer, none of the correct APIs are ranked by the TF-IDF-based matcher as its first 5 choices. Worse, nearly half of those are ranked above 100, making the synthesis time almost prohibitively long. 
We believe this is because the lexical overlap between the names of similar APIs in those two libraries is much smaller compared to the deep learning migration task. For example, dplyr's \lt{arrange} and panda's \lt{sort\_values} provide the same functionality (they both sort the rows by a given column), but the function names are different. In this way, Tfidf-GloVe can take advantage of the pretrained embeddings to explore the similarities between APIs beyond simple TF-IDF matching.

In Figure \ref{fig:results_dplyr}, we show the time it takes to migrate each of the $20$ benchmarks with a timeout of $3600$ seconds when using word embeddings. We solve $18$ out of $20$ collected benchmarks in under $102.5$ seconds. The average run time for $18$ benchmarks is $17.31 \pm    22.59$ seconds and a median of $12.19$ seconds. 
Note that for this task we did not consider error messages, nor specifications since we wanted to test how a basic version of \method would behave in a new domain. Moreover, for this domain, all the refactored benchmarks only used one-to-one mappings since no additional reshaping was needed before invoking \pandas APIs. Even with these conditions, we show that we are able to successfully refactor code for a new domain across different languages.

\begin{figure}[t]
    \centering
    \begin{tikzpicture}[scale=0.95]
        \begin{axis}[    xmode=linear,  ymode=log,
        width=6.3cm, height=2.6cm, grid=major,
        scaled y ticks = false,
        %yticklabel style = {/pgf/number format/fixed,
        %                     /pgf/number format/precision=3},
        tick label style = {font=\footnotesize},
        scale only axis,
        xtick={1, 5, 10, 15, 20},
        legend style={at={(0.55,0.85)}},
        xmax = 20 ,ymax = 10000, ymin = 1, xmin=1,
        xlabel = {\footnotesize{Instances solved}},
        ylabel = {\footnotesize{Time (s)}},
        ]

        \addplot+[draw=red!90, mark = x, fill opacity=1, draw opacity = 1, mark size=2] table [x index=0, y index=1, col sep=comma] {dplyr_pd.csv};
        %\addlegendentry{TF-IDF};
        \addplot [black, dotted, thick] coordinates {(0,3600) (20,3600)}
            node [below, xshift=-2.5cm]{\scriptsize{$3600$ seconds timeout}};

    \end{axis}
    \end{tikzpicture}
    \caption{Execution time for each benchmark of the dplyr-to-pandas task with a timeout of 3600 seconds.} %Tfidf-GloVe is used as the API matching method and no specification or error message is used for this task.}
    \label{fig:results_dplyr}
    \vspace{-2mm}
\end{figure}
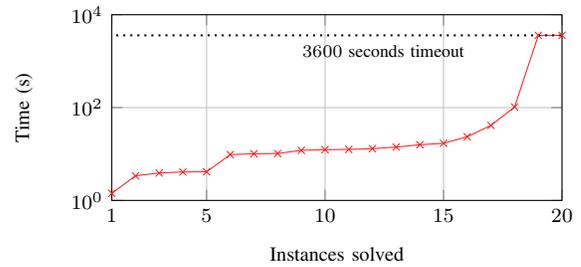

\section{Limitations and Discussion}
\label{sec:Limitations}
Here we discuss the main limitations of our method and possible challenges for
extending \method's ability to refactor new APIs, even potentially beyond the
domain of data science. 

\vspace{1ex}
\noindent\textbf{Benchmarks.} 
Our evaluation of \method uses benchmarks from well-known deep learning
tutorials and architectures. However, they are all feed-forward networks,
effectively sequences of API calls where  the output of the current layer is the
input of the next layer. There may be more applications that share this feature, but
support for more complex structure is likely necessary to adapt to other domains.
Additionally, and naturally, the APIs in the benchmarks we collected may be biased and not reflect the set of APIs developers actually use.

To assess this risk, 
we checked the degree to which the APIs used in our benchmarks
appear to be widely used on other open-source repositories on GitHub. 
To do this, we collected the top 1015 starred repositories
that have \tf as a topic tag, which contains over 8 million lines of code and over 500K \tf API calls. We found that 76\% of the 1000+ repositories use API calls included in our benchmarks at least once, which validates some representativeness of our collected benchmarks.

\vspace{1ex}
\noindent\textbf{Automatic testability.} One benefit of the data science/scientific
computing domain is that much of the input, output, and underlying methods are
typically well-defined.  As a result, it is particularly easy to test and verify
the correctness of individually migrated calls, which can be processed in
sequence. There may be other types of libraries that share these types of
characteristics, like string manipulation or image processing libraries, whose intermediate outputs are strings/images.
We also assume
user-provided tests. Given the migration task, it is reasonable to assume the user has 
tests (the code must be sufficiently mature to justify migrating, after all),
but a more general solution might benefit from automatically generating tests,
which would both alleviate the input burden on the user and, potentially, reduce
the risks of overfitting.  In our current implementation, we moreover use the
provided tests to construct smaller test cases for each
mapping. This is
particularly easy in this domain, because data science and deep learning
API calls are often functional in their paradigm.  Adapting the technique to
other paradigms would require more complex test slicing or
generation to support synthesis.

\vspace{1ex}
\noindent\textbf{API Matching.} \reb{Using the GloVe model for the API Matching often results in out of vocabulary problems because of API names and descriptions often use of data science specific terminologies, especially acronyms and abbreviations (e.g. ``conv2D'' for 2-dimensional convolution, ``LSTM'' for long-short-term-memory). The GloVe model is not trained for this domain, and the out of vocabulary problems explain the limited success of the Tfidf-GloVe model when compared to a plain TF-IDF. To address this problem, it would be necessary to train a new set of embeddings with focus on data-science jargon, which is out of scope.}

\vspace{1ex}
\noindent\textbf{Error message understanding.} 
%\an{I trimmed down this part}
The error message understanding model is built on four domain specific
lexico-syntatic patterns, which we identify as hyponyms when they appear in an error message. 
We propose the hyponyms based on the specific syntax of DL API error messages, thus take non-trivial human effort to make it generalize to error messages that appear when calling APIs from libraries of other domains. However, we believe the idea of program mutation (Step 3 of Fig. \ref{z3_gen}) is still widely applicable for the purpose of generating SMT constraints when dealing with error messages.

\vspace{1ex}
\noindent \textbf{Synthesis.} Our approach supports one-to-many mappings but it restricts the mapping to \emph{one} API of the target library and one or more reshaping APIs. However, this could be extended to include \emph{many} APIs of the target library at the cost of slower synthesis times. An additional challenge is to support many-to-one or many-to-many mappings since this would require extending our synthesis algorithm. However, even with the current limitations, our experimental results show that the current approach can solve a diverse number of benchmarks. 
%which  its applicability to data science tasks.

\vspace{1ex}
{\noindent \textbf{Generalizability.} } \reb{SOAR applies best to well-documented APIs with easily decomposable tests (i.e., calls have well-defined semantics and limited side effects). Deep Learning and Data Science APIs have these properties, and are popular, rapidly evolving, and used by programmers with a variety of backgrounds. We focus on them in the interest of impact. SOAR likely generalizes easily to domains that share these properties, like string or image manipulation libraries. Nonetheless, SOAR always requires a one-time effort to be instantiated in any domain. Specifically, SOAR needs: (1) a crawler and a parser to collect documentation used to build the API matching model and specification constraints; this step can be facilitated with tools like python's built-in function \textit{help} if API's are well-documented; (2) an error message message understanding model (which can be simply based on phrase structure rules). We do not study the effort needed to provide these two requirements; however, we believe it is significantly lower than building a static migration tool from scratch.}

\vspace{1ex}
\noindent\textbf{Correctness.} 
Since we evaluate our migration tasks using test cases, it is always possible
for our approach to overfit to these tests. However, this threat can be
mitigated if the user provides a sufficiently robust test set that provides
enough coverage.
Additionally, code written to
different APIs may be functionally equivalent, but demonstrate different
performance characteristics, which we do not evaluate.  However, this fact is
one reason users might find \method useful in the first place: a desire to
migrate code from one library to another that is more performant for the given
use case.

\vspace{1em}
Overall, we focus our design and evaluation on deep learning and data science libraries.
These libraries have properties that render them well-suited to our task in
terms of common programming paradigms, and norms, such as in the API
documentation.  However, we believe this is also a particularly useful domain to
support, given the field's popularity and how quickly it moves, how often new libraries are
released or updated, as well as the wide variety of skill sets and backgrounds
present in the developers who write data science or deep learning code. Automation
of migration and refactoring in this domain is very minimal, 
and we design \method as a step towards better
tool support for this diverse and highly active developer population.

\section{Related Work}
\label{sec:Background}

\subsection{Automatic Migration}
Existing work on automatic API migration uses example-based migration techniques.
Lamothe \textit{et al.} \cite{lamothe2018a4} proposed an approach that automatically learns API migration patterns using code examples and identified 83 API migration patterns out of 125 distinct Android APIs. Fazzini \textit{et al.} \cite{fazziniapimigrator} proposed APIMigrator, which learns from how developers from existing apps migrate APIs and uses differential testing to check validity of the migration. They were able to achieve 85\% of the API usages in 15 apps, and validated 68\% of those migrations. 
Meditor~\cite{xu2019meditor} mines open source repositories and extracts migration related code changes to automatically migrate APIs. Meditor was able to correctly migrate 218 out of 225 test cases. Unlike prior API migration tools, SOAR can migrate code without existing code examples.

%Xu \textit{et al.} \cite{xu2019meditor} introduced Meditor, which mined open source repositories and extracted migration related code changes to automatically migrate APIs. Meditor was able to correctly migrate 218 out of 225 test cases. Unlike prior API migration tools, SOAR can migrate code without existing code examples. 

SOAR also relates to automatic migration on APIs between different programming languages. Zhong \textit{et al.} proposed MAM \cite{zhong2010mining} and mined 25,805 unique API mapping relations of APIs between Java and C\# with 80\% accuracy.
Nguyen \textit{et al.} proposed StaMiner \cite{nguyen2014statistical}, which is a data-driven approach that statistically learns the mappings of APIs between Java and C\#. Bui \textit{et al.} \cite{bui2019sar} used a large sets of programs as input and generated numeric vector representations of the programs to adapt generative adversarial networks (GAN). Bui \textit{et al.} then identified the cross-language API mappings via nearest-neighbors queries in the aligned vector spaces. 
Again these methods largely rely on existing training data, such as MAM and StaMiner \cite{zhong2010mining,nguyen2014statistical} mine mappings from parallel equivalent code from two languages (Java and C\#), where \method only leverages the documentation for migration.

\subsection{Program Synthesis}

Program synthesis has been used to automate tasks in many different domains, such as, string 
manipulations~\cite{DBLP:conf/icse/DesaiGHJKMRR16}, table transformations~\cite{DBLP:conf/pldi/FengMGDC17}, SQL queries~\cite{DBLP:journals/pacmpl/Yaghmazadeh0DD17}, and synthesis of 
Java functions~\cite{DBLP:journals/pacmpl/ShiSL19}. However, its usage for program 
refactoring is scarce. ReSynth~\cite{DBLP:conf/oopsla/RaychevSSV13} uses program synthesis 
for refactoring of Java code by providing an interactive environment to programmers, where they indicate the desired transformation with examples of changes.
% Next, the program synthesizer synthesizes a sequence of refactorings that 
% include the edits made by the programmer. To prune the search space, ReSynth only considers 
% program entities that were edited by the user and uses A$^*$ search to synthesize the 
% refactored code. 
Our approach differs from ReSynth since we do not require the user to 
provide a partially refactored code. Since our problem domain is API migration, it is 
unlikely that the user knows all the required APIs from the target library and can perform 
these edits. 
% However, if this was the case, then our approach could still be used by having 
% the user provide the program sketch and have the program synthesizer fill the holes with the 
% correct parameters. 
% \an{I don't think we want to spend a paragraph on describing a prior work, unless it's super important?}

NLP can be used to synthesize programs directly from natural language~\cite{DBLP:conf/icse/DesaiGHJKMRR16,DBLP:journals/pacmpl/Yaghmazadeh0DD17} or to guide the search of the program 
synthesizer~\cite{DBLP:conf/pldi/Chen0YDD20,DBLP:conf/sigsoft/ChenMF19}. For instance, NLP 
has been used to synthesize tasks related to repetitive text editing~\cite{DBLP:conf/icse/DesaiGHJKMRR16}, SQL queries~\cite{DBLP:journals/pacmpl/Yaghmazadeh0DD17}, and synthesis of 
regular expressions~\cite{DBLP:conf/pldi/Chen0YDD20}. One can also combine input-output 
examples with a user-provided natural description to have a stronger specification and 
achieve better performance~\cite{DBLP:conf/pldi/Chen0YDD20,DBLP:conf/sigsoft/ChenMF19}.
Our approach follows this trend of work where we combine NLP to guide the program 
synthesizer with input-output examples that provide stronger guarantees in the synthesized 
code. However, instead of using a natural description provided by the user, our approach 
uses documentation from libraries to guide the search.

% error messages
Using error messages from the compiler or interpreter is not common in program 
synthesis. The most relevant approach to ours is the one from Guo \emph{et al.}~\cite{DBLP:journals/pacmpl/GuoJJZWJP20} where they use type error information to refine 
polymorphic types when synthesizing Haskel code. In contrast, \method uses error messages 
from the interpreter not to refine the type information but to restrict the domain of the 
parameters and to prune the search space.

\reb{Finally, our synthesis strategy is based on program sketching and program enumeration. This approach has close parallels (e.g., \cite{DBLP:conf/popl/FengM0DR17, DBLP:conf/pldi/WangCB17}) and is extremely common in modern synthesizers because it provides a simple way of splitting the search space. Our approach can also be seen a generate-and-validate strategy using test-cases as an oracle to evaluate migration success, which is also widely used repair engines \cite{DBLP:journals/cacm/GouesPR19}.}

\section{Conclusions}
\label{sec:Conclusions}
API selection and maintenance is an important and difficult task for software
development. To match evolving software, developers often have to manually refactor APIs, which is a tedious and error-prone job.
We proposed \method to take advantage of API documentation and error messages as a rich sources of information intended for developers.
It uses natural language processing and
program synthesis to automatically write refactored API calls. It is
particularly well-suited for data science or deep learning library refactoring,
a prevalent use case in modern development where tool support is positioned to
have particular impact.  \method collects
information from both API documentation and error messages to generate logical
constraints that can be used to limit the synthesizer search space. Unlike prior
approaches to automatic API migration, \method requires no training data, and
its output is guaranteed to compile and pass existing tests. Our empirical
evaluation shows that \method can successfully refactor 16/20 of our benchmarks
for the deep learning domain with an average time of 97.23 seconds, and 18/20 of the benchmark
set for data wrangling tasks with an average time of 17.31 seconds.

\section*{Acknowledgments} 
This work was partially supported under National Science Foundation Grant Nos. CCF-1910067, CCF-1750116 and CCF-1762363, and by Portuguese national funds through FCT, Fundação para a Ciência e a Tecnologia, under PhD grant SFRH/BD/150688/2020 and projects UIDB/50021/2020, DS\-AIPA/AI/0044/2018, and project ANI 045917 funded by FEDER and FCT. All statements are those of the authors, and do not
necessarily reflect the views of any funding agency.

\bibliographystyle{ieeetr}
\bibliography{refs}

\end{document}